\documentclass[twocolumn]{aa}
\usepackage{graphicx,epsf,psfig,epsfig,txfonts}

\def\integral{{\it INTEGRAL}}
\def\vlt{{\it VLT}}
\def\rxte{{\it RXTE}}
\def\chandra{{\it Chandra}}
\def\1e{{1E~1740.7--2942}}

\newcommand{\g}{$\gamma$}
\newcommand{\gro}{{\it CGRO}}

\begin{document}

   \title{1E 1740.7--2942: temporal and spectral evolution from INTEGRAL and RXTE 
observations\thanks{Based on observations with \integral, an ESA project with instruments and science data centre funded by ESA member states (especially the PI countries: Denmark, France, Germany, Italy, Switzerland, Spain), Czech Republic and Poland, and with participation of Russia and the USA.}}

   \author{M. Del Santo\inst{1}, A. Bazzano\inst{1}, A. A. Zdziarski\inst{2}, D. M. Smith\inst{3}, 
N. Bezayiff\inst{3}, R. Farinelli\inst{4},G. De Cesare\inst{1}, P. Ubertini\inst{1}, 
A.J. Bird\inst{5}, M. Cadolle Bel\inst{7}, F. Capitanio\inst{1}, A. Goldwurm\inst{7},
A. Malizia\inst{6}, I. F. Mirabel\inst{7}, L. Natalucci\inst{1}, C. Winkler\inst{8} 
}

\offprints{Melania Del Santo, delsanto@rm.iasf.cnr.it}

\institute{ Istituto di Astrofisica Spaziale e Fisica cosmica - CNR, via del Fosso del Cavaliere 100, 00133 Roma, Italy
\and Nicolaus Copernicus Astronomical Center, Bartycka 18, 00-716 Warszawa, Poland
\and Department of Physics, University of California, Santa Cruz, Santa Cruz, CA 95064
\and Dipartimento di Fisica, University of Ferrara, via del Paradiso 12, 44100 Ferrara, Italy
\and School of Physics and Astronomy, University of Southampton, SO17 1BJ, UK
\and IASF/CNR, sezione di Bologna, via P. Gobetti 101, 40129 Bologna, Italy     
\and  CEA Saclay, DSM/DAPNIA/SAp (CNRS FRE 2591), F-91191 Gif sur Yvette Cedex, France
\and Research and Scientific Support Department of ESA, ESTEC, Postbus 299, 2200 AG Noordwijk, The Netherlands
        } 

\authorrunning{M. Del Santo et al.}
\titlerunning{1E 1740.7--2942: temporal and spectral evolution}
 
\date{Received... accepted...}

\abstract{We present results of the monitoring of the black hole candidate \1e\ with \integral, 
in combination with simultaneous observations by \rxte. We concentrate on broad-band spectra from 
\integral/IBIS and \rxte/PCA instruments. During our observations, the source spent most of 
its time in the canonical low/hard state with the measured flux variation within a factor of two. 
In 2003 September the flux started to decline and in 2004 February it was below 
the sensitivity level of the \integral\/ and \rxte\/ instruments. 
Notably, during the decline phase the spectrum changed, becoming soft and typical of black-hole 
binaries in the intermediate/soft state. 
}

\maketitle

\begin{table*}
\caption{\integral\/ observations of \1e.}
\begin{center}
\begin{tabular}{c c c c c}  
Data Set & Observation Period & \integral\ JD$^{\rm a}$ & Revolutions & Exposure (ks) \\
\hline
GCDE 1$\backslash$a  & 2003/02/28--2003/04/23       &  1166.4--1205.6 & 50--64 &    {690} \\ 
GCDE 1$\backslash$b  &  2003/08/10--2003/10/14      & 1317.7--1382.2  & 100--122 &   {720} \\
XTE J1720--318 ToO & 2003/02/28--2003/03/02 & 1154.2--1156.3 & 46   &    {168} \\
H1743--322 ToO  &  2003/04/06--2003/04/22 & 1191.6--1207.5 & 58, 61, 63 & {280} \\
GCDE 2$\backslash$a &  2004/02/16--2004/04/20  &  1507.7--1571.5 &  164--185  & 930 \\ 
\hline
\end{tabular}
\end{center}
\small{$^{\rm a}$ The \integral\ Julian Date (IJD) starts from 2000 January 1 and it corresponds to ${\rm MJD}={\rm IJD}+51544$. }
\label{log}
\end{table*}

\begin{figure*}
\centering
\includegraphics[height=15cm,angle=90]{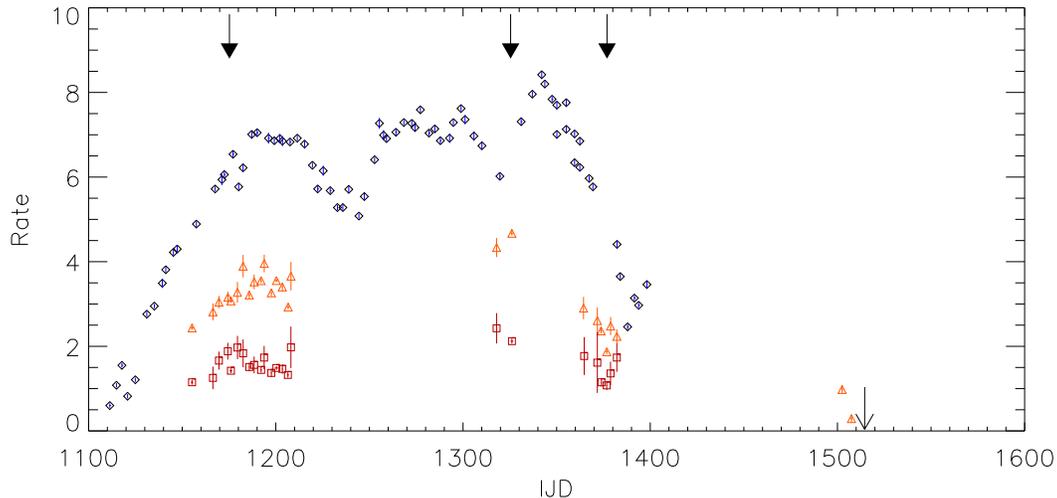}
\caption{Lightcurves from PCA (8--25 keV, blue diamonds), and from IBIS/ISGRI in the two bands 30--50 keV (orange triangles)
and 50--80 keV (red squares). Solid arrows indicate periods used for the spectral analysis. The differences in the sizes 
of the ISGRI error bars are due to the different exposure times. 
For the clarity of display, the lightcurve for the 50--80 keV band has been multiplied by 0.6. 
Simple arrow indicates when the source started to be not visible by IBIS, anymore.}
\centering
\label{fig:xte}
\end{figure*}

\section{Introduction} 

The X-ray source \1e was discovered in 1984 by the {\it{EINSTEIN}\/} observatory 
(\cite{hertz84}) and its hard X-ray emitting nature was firstly reported 
by \cite{skinner87}. Based on the spectral shape of its soft \g-ray 
emission and on the similarities with Cyg X-1, it was classified as a black hole 
candidate (BHC) and reported as the strongest persistent source in the 
Galactic Centre region (\cite{sunyaev91a}).

After the discovery of a double-sided radio-emitting jet, \1e\ was classified as a microquasar near 
the Galactic Centre by \cite{mirabel92}. In 1991, it was also suggested as a possible positron-electron 
annihilation source due to a {\it{GRANAT}}/SIGMA observation of a high energy feature (\cite{bouchet91}; 
\cite{sunyaev91b}) in the spectrum, though its definitive confirmation has not been reported so far. 
In fact, a simultaneous observation by \gro/OSSE (\cite{jung95}) has severely constrained the suggested 
feature intensity with an upper limit substantially below the flux measured by SIGMA (\cite{cordier93}).

Most of the time, the source is observed in the typical BHC low/hard state, 
characterised in the X-ray domain by an absorbed power law with the photon index 
of $\Gamma\sim 1.4$--1.5 (\cite{gallo02}) and a cutoff at high energies 
(\cite{sidoli99}). The high hydrogen column density toward the source, 
recently estimated by \chandra\/ as $N_{\rm H}\simeq 1.05 \times 10^{23}$ cm$^{-2}$ 
(\cite{gallo02}), does not permit any identification of its optical counterpart.

During \rxte\/ observations of \1e, an interesting correlation between the 
time derivative of the flux and the hardness was discovered; 
on the other hand, no such correlation was found in Cyg X-1 (\cite{smith_a}). A periodic 
modulation with amplitude 3--4\% at $12.73 \pm 0.05$ days has been measured and 
interpreted as the orbital period, suggesting that the object could have a 
red-giant companion (\cite{smith_b}). 
On the contrary, \cite{marti00} reported on the upper limit 
for an IR counterpart with \vlt\/ which seems to exclude a red-giant companion.

Preliminary \integral\/ results have been already presented in \cite{delsanto04}; in this paper 
it was reported on a flux variability of a factor of 2.5 and on the absence of any correlation
between photon index and 20--80 keV flux during the low/hard state period. 
No spectral variability was presented, since the analysis was not yet completed at that time.

In the present work, we report on one year of almost simultaneous \integral\/ and \rxte\/ 
observations and on the first period of \integral\/ second year observations. 
Details of the data set and scientific analysis are presented in 
Section 2. Temporal and spectral behaviour are reported in Section 3 and 
discussed in Section 4.

\section{Observations and data analysis}

We analyse here \integral\/ observations of \1e\ performed in the framework of 
the Core Programme. This includes roughly 800 science windows (ScWs) collected during the 
first year of the Galactic Centre Deep Exposure (GCDE). The total exposure of 
1.4 Ms consists of two parts separated by $\sim$4 months. 
The second year of the 
GCDE has not been completed yet; 930 ks have been performed so far and they are 
included in this work.

We have also added some pointings of the Targets of Opportunity (ToO) 
observations of XTE J1720--318 (\cite{cadolle04}) and H1743--322 (\cite{parmar03}), corresponding 
to the new \integral\/ source IGR J17464--3213 (\cite{revniv03}), which 
included \1e\ in the field of view. In Table \ref{log}, we give the log of the 
observations. The \integral\/ Core Programme pointings have a 
duration of 1800 s each for the GCDE and 2200 s  each for the ToOs (see 
\cite{winkler99} for Core Programme details).

\rxte\/ has also monitored \1e\ since its first week of science operations in 
1996. The monitoring began as monthly pointings of $\sim$1000--1500 s, the
frequency of which has increased over the years to about twice per week now. The \rxte\/ 
observations simultaneous with those by \integral\/ cover three different 
periods: 2003 March 13--April 26 with the total exposure of 
$\sim$19 ks; 2003 August 12--24 with an exposure of $\sim$5 ks; and 
the third one, which consists of only two pointings at the beginning of 2003 
October. 

We use here data collected with the ISGRI (\cite{lebrun03}) detector of the imager IBIS 
(\cite{ubertini03}) on board \integral\/ for the soft \g-ray range, and with the 
\rxte/PCA detector for the X-rays. The IBIS Partially Coded Field Of View (PCFOV) is 
$29\degr \times 29\degr$ at the zero response, but the full instrument 
sensitivity is achieved in the $9\degr \times 9\degr$ Fully Coded Field of View 
(FCFOV). For the temporal analysis, we took into account all the data, whereas 
for the spectral extraction, we selected only pointings with the source  
within the FCFOV, for which a reliable detector spectral response matrix is available.

Raw data corresponding to each pointing have been pre-processed and organised in 
ScWs by the Integral Science Data Centre (\cite{courvoisier03}) pipeline. The 
IBIS scientific data analysis software (\cite{goldwurm03}) used is included in 
the \integral\/ off-line analysis software (OSA) release 3.0. The lightcurves 
have been extracted by the images in two energy bands: 30--50~keV and
50--80~keV. The ISGRI spectra have been extracted in 16 
logarithmic bins and fitted using the response matrix corrected and
tested by the IBIS team, and now delivered with the OSA 4.0 release.

The PCA data were analysed with FTOOLS v.\ 5.2 of the NASA High Energy 
Astrophysics Science Archive Research Center.  Due to the crowded field near the 
Galactic Centre, the pointing axis was offset by $\ga 0.5\degr$ to avoid nearby 
sources. To eliminate the spectral contribution from the Galactic diffuse 
emission as exactly as possible, a background field was taken 
at a position symmetrical about the Galactic Centre from 
the source field.  A map of the source and background fields with nearby sources 
is given in \cite{main99}.

Spectral fitting have been performed with XSPEC v.\ 11.3.1 and single parameters 
uncertainties have been calculated at the 90\% confidence level, i.e., for 
$\Delta \chi^2=2.71$. Systematic errors of 2\% for the ISGRI spectra (which 
approximately corresponds to the current accuracy of the calibration) and 1\% 
for the PCA spectra have been added in quadrature.

\section{Results}
During the GCDE 1$\backslash$a (see Table \ref{log}), the 30--50~keV 
ISGRI count rate increased by 50\%. We measured the rate of $2.43 \pm 0.04$ 
s$^{-1}$ at the beginning of 2003 March and $3.66 \pm 0.33$ s$^{-1}$ after one 
month; these values correspond to $\sim$37 and $\sim$56~mCrab, respectively. 
After four months, we observed a further increase up to 70~mCrab in the rev.\ 
103 (2003 August), in the same energy range. In 2003 September, the flux began 
to decline, and at the end of the GCDE 2003 it reached 25~mCrab.

In Fig. \ref{fig:xte}, we show the PCA lightcurve (8--25~keV) together with two 
ISGRI lightcurves in the 30--50~keV and 50--80~keV energy ranges. 
Similar temporal behaviours appear in all three energy bands.

In 2004, when \integral\/ pointed again at the Galactic Centre, the \1e\ flux was 
dramatically lower. The source was detected at $\sim$10 mCrab during the rev.\ 
164 and $\sim$3 mCrab during the rev.\ 165 (2004 February, IJD $\sim$1500). 
Soon after (simple arrow in Fig. \ref{fig:xte}) no detection was possible both 
with IBIS in \g-rays (also reported by \cite{grebenev04})
and with PCA in the X-ray domain (D. M. Smith, private communication). 
Such a behaviour is not unusual, even though rare; in
spring 1991 the source flux went down to the {\it GRANAT}/SIGMA sensitivity 
level in the 40--150 keV range (\cite{cordier94}).

In order to study the source spectral evolution, we have chosen three periods 
corresponding to different flux levels, namely the rev.\ 53, 103 and 119--120, 
indicated by the solid arrows in Fig.\ \ref{fig:xte}. We initially fitted power law 
and an e-folded power law models. For the rev.\ 53 and 103 we obtained 
the following best fit parameters with the $N_{\rm H}$ fixed at $1.05\times 10^{23}$ cm$^{-2}$ 
(using the model \texttt{WABS} in XSPEC): the photon index $\Gamma=1.59 \pm 0.03$ and the 
e-folding energy $E_{\rm c}=146_{-32}^{+53}$ keV with $\chi_{\nu}^2=63/54$
for the first period, and $\Gamma=1.34 \pm 0.03$, $E_{\rm c}=90_{-13}^{+18}$ 
keV with $\chi_{\nu}^2=63/54$ for the second one. 
The mean spectrum of the revolutions 119 and 120 has been fitted 
by a simple power-law with $\Gamma=2.25 \pm 0.05$, which gives $\chi_{\nu}^2=82/51$.

\begin{figure}
\centering
\includegraphics[width=8.5cm]{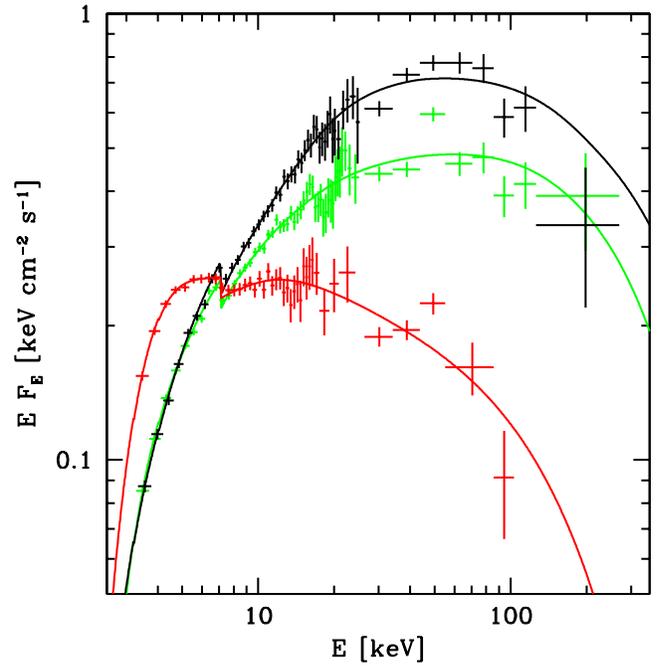}
\caption{The observed (absorbed) spectra, as fitted by the thermal Comptonization model, \texttt{COMPPS}. 
The green, black and red 
spectra correspond to the rev.\ 53 (hard state), 103 (hard state) and 119--120 (soft state), respectively.}
\centering
\label{fig:spectra_ab}
\end{figure}

\begin{figure}
\centering
\includegraphics[width=8.5cm]{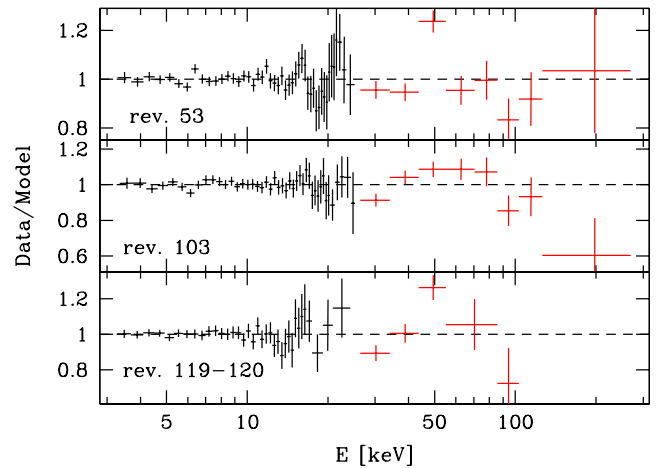}
\caption{Residuals for the three data set.}
\centering
\label{fig:residual}
\end{figure}

Then, two thermal Comptonization models have been used: \texttt{COMPTT} of 
\cite{titarchuk94} and \texttt{COMPPS}\footnote{Model available at 
ftp://ftp.astro.su.se/pub/juri/XSPEC/COMPPS.} of \cite{poutanen96}. 

\texttt{COMPPS} is a highly accurate iterative scattering Comptonization model, in 
which subsequent photon scatterings are directly followed (see, e.g., \cite{zdz00}; \cite{zdz03}): 
this model has been extensively tested against Monte Carlo results.
On the other hand, \texttt{COMPTT} is based on an approximate solution of the kinetic equation 
with some relativistic corrections, and the resulting spectra are only 
approximate (see, e.g., discussion in the appendix of Zdziarski, Johnson \& Magdziarz (1996)). 
Unlike \texttt{COMPTT} where the input spectrum of low--energy photons is a Wien law, 
\texttt{COMPPS} assumes for the seed photons a simple blackbody or a multicolor disc blackbody spectrum.

We have allowed a free column density, but constrained it to $N_{\rm H}\geq 1\times 
10^{23}$ cm$^{-2}$ (Gallo \& Fender 2002). With the former we have obtained a 
relatively low temperatures and high Thomson optical depths for the first 
two spectra, namely $kT_{\rm e}=26_{-3}^{+6}$ keV, $\tau=1.9_{-0.2}^{+0.3}$, 
$\chi_{\nu}^2=59/54$, and $kT_{\rm e}=20 \pm 2$ keV, $\tau=2.9 \pm 0.1$, 
$\chi_{\nu}^2=55/53$, respectively. We have not been able to constrain the 
temperature for the third spectrum.

\begin{table*}
      \caption{The parameters of the PCA/ISGRI spectra fitted with the {\texttt COMPPS} model. }
        \centering  
         \label{parameters}
     $$   
         \begin{array}{lccccccc}
            \hline
            \noalign{\smallskip}
{\rm Rev.} & {N_{\rm H}}^{\mathrm{a}} & kT_{\rm bb} & kT_{\rm e} &  \tau & \Omega/ 2\pi & {F_{\rm bol}}^{\mathrm b} & \chi^2_\nu \\
            \noalign{\smallskip}
& 10^{22} {\rm cm}^{-2} & {\rm keV} & {\rm keV} &&& {\rm erg\, cm}^{-2} {\rm s}^{-1}\\
            \noalign{\smallskip}
            \hline
            \noalign{\smallskip}
53 & 12_{-2}^{+2} & 0.50^{+0.06}_{-0.11} & 110_{-30}^{+20} &  1.1_{-0.3}^{+0.6} & 0.3_{-0.2}^{+0.2}& 3.6\times 10^{-9} & 70/51\\
             \noalign{\smallskip}
103 & 11_{-1}^{+2} & 0.41^{+0.09}_{-0.10} & 150_{-50}^{+60} &  0.9_{-0.4}^{+0.4} & 0.9_{-0.2}^{+0.3}& 4.8\times 10^{-9} & 50/51\\
             \noalign{\smallskip}
119\!\!-\!\!120 & 10^{+1} & 0.68^{+0.06}_{-0.05} & 70_{-8}^{+11} &  1.0_{-0.2}^{+0.2} & 0.0^{+0.1}& 2.1\times 10^{-9} & 44/47\\
           \hline
         \end{array}
     $$
\begin{list}{}{}
\item[$^{\mathrm{a}}$] Constrained to $\geq 10^{23}$ cm$^{-2}$.
\item[$^{\mathrm{b}}$] The bolometric flux of the unabsorbed best-fit model spectrum.
\end{list}
\end{table*}

\begin{figure}
\centering
\includegraphics[width=8.5cm]{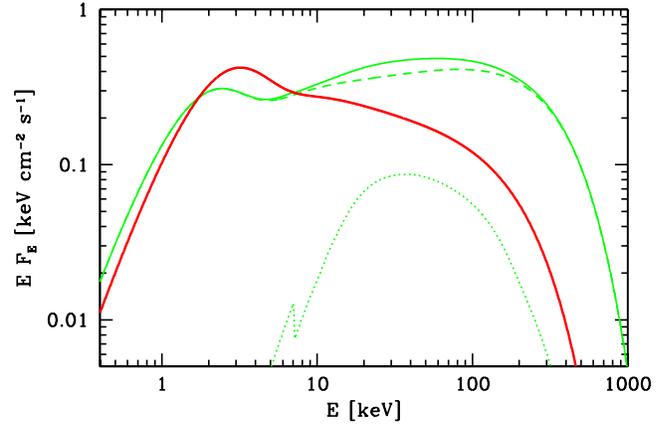}
\caption{The unabsorbed thermal Comptonization for the rev.\ 53 (green) and 119--120 (red, heavy). 
For the first spectrum, its components due to thermal Comptonization of blackbody photons 
(dashed curve) and Compton reflection (dotted curve) are shown.
}
\centering
\label{spectra_unabs}
\end{figure}

The fit results with the \texttt{COMPPS} are given in Table~\ref{parameters}. 
We chose a geometry where Compton scattering is computed by mean of
approximate treatment of radiative transfer using escape probability for a sphere.
The model also includes Compton reflection (\cite{magd95}), which is found to be 
important for the rev.\ 103 (F-test probability of $10^{-7}$).   
We also fixed the ionization parameter at 0 as PCA has poor energy resolution.
The three absorbed spectra are plotted in Fig.\ 
\ref{fig:spectra_ab} and in Fig.\ \ref{fig:residual} the related residuals are shown. 
Fig.\ \ref{spectra_unabs} shows the intrinsic, unabsorbed spectra for the rev.\ 53 and 119--120. 

We note that our PCA data have been background-subtracted in a non-standard way (Section 2). 
This procedure is supposed to remove, in particular, the Galactic ridge Fe K$\alpha$ emission. 
In our PCA data, we see no evidence of any Fe K line; however, this may be simply due 
to residual inaccuracies of the background subtraction.

Nearby sources are excluded by placing the centre of the 1$^\circ$
field of view about 0.5$^\circ$ away from \mbox{1E 1740.7-2942},
and the strong Galactic diffuse background is subtracted
with the help of pointings to an empty field symmetrically
positioned across the GC.

The spectra indicate an apparent state transition occurring from the rev.\ 103 to 119--120. The 
spectra for rev.\ 53 and 103 are consistent with those typical for black-hole 
binaries in the hard state whereas that for rev.\ 119--120 is more typical 
for the soft or intermediate state (\cite{zdz04a}). 
Surprisingly, the Compton reflection component appears the weakest in the last 
data set, which is not typical for black-hole binaries (\cite{gier99}).

\section{Discussion and conclusions}

We have observed several variability patterns of \1e. First the estimated bolometric 
flux increased by $\sim$1.5 in 6 months (2003 February--August) with an almost 
constant spectral shape. The spectrum, with a hard X-ray index, a Compton-reflection 
component, and a cutoff corresponding to $kT_{\rm e}\sim 100$ keV, is typical for 
the low/hard state of the black-hole binaries, e.g., Cyg X-1 (\cite{gier97}) 
or GX 339--4 (\cite{ward02}).  

Then, however, we have observed a spectral state transition. In 2003 October, the flux in 
the low-energy part of the spectrum has increased, and our fits show an increase of the 
disc blackbody temperature from 0.4--0.5 keV in the hard state up to $\sim$0.7 keV. 
The power law has substantially softened, with $\Gamma$ increasing from 1.3--1.5 up to $\sim$2.3. 
These spectral characteristic are peculiar of the soft state, e.g., of Cyg X-1 (\cite{gier99}). 

Interestingly, the flux of this state was significantly lower than that of the hard state. 
Although the opposite behaviour, with the flux in the soft state larger than that 
in the hard state, is most typical for black-hole binaries, the behaviour observed 
by us is also seen in other low-mass X-ray binaries, e.g., in GX 339--4.
Smith et al. (2002a) have often observed also GRS 1758-258 at low luminosities in the typical soft state.
This is due to the existence of two accretion solutions at a given accretion rate, which allows 
for such a hysteretic behaviour (e.g., \cite{zdz04b}). This transient-like 
type of behaviour is consistent with occasional quenching of the X-ray emission of \1e, 
as also confirmed by our study. 

Assuming a distance of 8.5 kpc and 10 $M_{\odot}$ for the compact object, 
we estimated the bolometric luminosity of \1e during rev. 53,
rev. 103 and rev 119-120 to be $\sim$0.01 $\it{L_{Edd}}$,
$\sim$0.03 $\it{L_{Edd}}$ and $\sim$0.02 $\it{L_{Edd}}$, respectively.
Our results are consistent with \cite{macca03}, who shows that the 
state transition luminosities are at about 1--4\% 
of the Eddington rate for most of the X-ray binaries.

An interesting feature of the observed soft state is the unusual weakness of the Compton 
reflection component, which usually becomes stronger in the soft state 
(e.g., \cite{gier99}) than in the hard state. We suggest that it 
is possible that the accretion disc in \1e\ precesses, and we could have caught 
the source in the moment when the disc is edge-on, just causing the reflection to be not observable.

Finally, we outline that in our spectra integrated over time scales of
20 ks and 100 ks, there is no evidence of any high energy feature.
In the energy range 485--535 keV in $\sim$1.4 Ms integration time, 
\cite{decesare04} report on an upper limit at $2\sigma$ level 
of roughly $2.0\times 10^{-4}$ ph cm$^{-2}$ s$^{-1}$, including systematic errors.  
Based on this result, the $2\sigma$ upper limits for our time scales of 20 ks and 100 ks
are $1.7\times 10^{-3}$ and  $7.4\times 10^{-4}$ ph cm$^{-2}$ s$^{-1}$, respectively.

\begin{acknowledgements}
This work has been supported by ASI via contract I/R/389/02 and I/R/041/02. 
MDS thanks Julien Malzac for precious scientific suggestions, Memmo Federici 
for the data archive support and Angelique Joinet for useful data analysis discussion. 
AAZ has been supported by KBN grants PBZ-KBN-054/P03/2001 and 1P03D01827.
We thank A. Parmar for providing access to proprietary data.
 \end{acknowledgements}

\end{document}